\documentclass[apj]{emulateapj-rtx4}

\shorttitle{}
\shortauthors{Kim et al.}
\begin{document}
\title{Accretion Rates of Red Quasars from the Hydrogen P$\beta$ line}

 \author{\textbf{Dohyeong Kim}\altaffilmark{1,2}, \textbf{Myungshin Im}\altaffilmark{1,2},
 \textbf{Eilat Glikman}\altaffilmark{3}, \textbf{Jong-Hak Woo}\altaffilmark{2}, and \textbf{Tanya Urrutia}\altaffilmark{4}}  

\altaffiltext{1}{Center for the Exploration of the Origin of the Universe (CEOU),
Astronomy Program, Department of Physics and Astronomy, Seoul National University, 
Shillim-Dong, Kwanak-Gu, Seoul 151-742, South Korea}
\altaffiltext{2}{Astronomy Program, Department of Physics and Astronomy, Seoul National University, Shillim-Dong, Kwanak-Gu, Seoul 151-742, South Korea}
\altaffiltext{3}{Department of Physics, Middlebury College, Middlebury, VT 05753, USA}
\altaffiltext{4}{Leibniz-Institut f\"{u}r Astrophysik Potsdam, An der Sternwarte 16, 14482, D-14482, Potsdam, Germany}

\email{dohyeong@astro.snu.ac.kr, mim@astro.snu.ac.kr}

\begin{abstract}
 Red quasars are thought to be an intermediate population between
 merger-driven star-forming galaxies in dust-enshrouded phase and normal quasars.
 If so, they are expected to have high accretion ratios, but their intrinsic dust extinction hampers reliable determination of Eddington ratios.
 Here, we compare the accretion rates of 16 red quasars at $z \sim 0.7$ to those of normal type 1 quasars at the same redshift range.
 The red quasars are selected by their red colors in optical through near-infrared (NIR) and radio detection.
 The accretion rates of the red quasars are derived from the P$\beta$ line in NIR spectra, which is obtained by the SpeX on the Infrared Telescope Facility (IRTF)
 in order to avoid the effects of dust extinction. 
 We find that the measured Eddington ratios ($L_{\rm bol}$/$L_{\rm Edd} \simeq 0.69$) of red quasars are significantly higher than those of normal type 1 quasars,
 which is consistent with a scenario in which red quasars are the intermediate population and the black holes
 of red quasars grow very rapidly during such a stage.
\end{abstract}

\section{INTRODUCTION}
 Most of our understanding of quasars is based on normal type 1 quasars
 found by X-ray, ultraviolet (UV), optical, and radio surveys \citep{grazian00,becker01,anderson03,croom04,risaliti05,schneider05,veron-cetty06,young09}.
 Moreover, normal type 1 quasars have been studied by using the luminosities in several wavelengths such as X-ray, UV, optical, and radio.
 However, several studies have suggested that the UV/optical quasar surveys could miss a large number of quasars
 because some quasars are obscured by intervening dust in their host galaxy \citep{webster95,cutri02}
 or by the interstellar medium of our galaxy \citep{im07,lee08}.

 These dust-obscured quasars would appear to have red colors, requiring different selection techniques to find them.
 Many studies have been carried out to search for red quasars
 (e.g., \citealt{webster95,benn98,cutri01,glikman07,glikman12,glikman13,urrutia09,banerji12,stern12,assef13,fynbo13,lacy13}).
 Such studies commonly take advantage of large area near-infrared (NIR) surveys such as
 the Two Micron All-Sky Survey (2MASS; \citealt{skrutskie06}), the UKIRT Infrared Deep Sky Survey (UKIDSS; \citealt{lawrence07}),
 the $\it{Spitzer}$ Wide-area Infrared Extragalactic Survey (SWIRE; \citealt{lonsdale03}),
 or the $\it Wide~field~Infrared~Survey~Explorer$ ($\it WISE$; \citealt{wright10}).
 In particular, red quasars are found by selecting for objects with very red colors in the optical through NIR based on
 2MASS \citep{cutri01,glikman07,glikman12,urrutia09}, UKIDSS \citep{banerji12,fynbo13,glikman13}, SWIRE \citep{lacy07,lacy13},
 or $\it WISE$ \citep{stern12,assef13}.
 In addition, X-ray surveys have been conducted to uncover dust-obscured quasars \citep{norman02,anderson03,risaliti05,young09,brusa10}.

 Pieces of observational evidence suggest that red quasars are a young population based on their
 (i) enhanced star formation activities \citep{georgakakis09};
 (ii) a high fraction of merging features \citep{urrutia08,glikman15}; (iii) young radio jets \citep{georgakakis12};
 and (iv) red continuum from dust extinction, possibly caused by their host galaxies \citep{glikman07,urrutia09}.
 These pieces of observational evidence point to a picture in which red quasars are the intermediate population between
 the merger-driven star-forming galaxies often seen as ultraluminous infrared galaxies (ULIRGs; \citealt{sanders88,sanders96})
 and normal blue quasars \citep{glikman07,glikman12,urrutia08,urrutia09,georgakakis09,georgakakis12}.
 In particular, simulations \citep{menci04,hopkins05,hopkins06,hopkins08} have shown that
 a major merger can trigger intense star formation and buried quasar activity.
 After that, the quasar grows with a high Eddington ratio but is still red because of remaining dust and gas,
 and the host galaxy is expected to show merger signatures and star formation activity.
 Finally, the quasar becomes a normal quasar after sweeping away the gas and dust from the quasar-driven wind.
 Such a scenario suggests that red quasars are in the intermediate stage between merger-driven star-forming galaxies
 and normal type 1 quasars.

 However, other explanations have been proposed to explain the existence of red quasars.
 \citet{wilkes02} suggested that the red colors of red quasars arise from
 the viewing angle in the quasar unification scheme when the accretion disk and the broad-line region (BLR) are viewed through a dust torus.
 Another suggestion is that red quasars have intrinsically red colors, i.e. an unusual covering factor of hot dust,
 a contamination from host galaxy light, or synchrotron emission that peaks at NIR wavelengths from radio jets
 \citep{puchnarewicz98,whiting01,rose14,ruiz14}.

 One observational test for red quasars as an intermediate population,
 under the merger-driven galaxy evolution scenario, is to investigate whether red quasars exhibit high accretion rates.
 In this regard, previous studies \citep{canalizo12,urrutia12,bongiorno14} have shown that
 red quasars could have higher accretion rates than normal quasars and different scaling relations of
 $M_{\rm BH}$--$L_{\rm B}$ and $M_{\rm BH}$--$M_{\rm \ast}$ compared to those of normal quasars.
 However, the accretion rates of red quasars are still controversial since most of the black hole (BH) mass and
 the bolometric luminosity estimators are based on flux measured in the UV or optical
 \citep{kaspi00,vestergaard02,mclure04,greene05}.

 A major difficulty in estimating BH masses using UV or optical light is that
 they are easily affected by dust extinction.
 For example, if a quasar is reddened by a color excess of $E(B-V)=2$ mag,
 its H$\beta$ and H$\alpha$ line fluxes are suppressed by factors of 860 and 110, respectively,
 assuming the Galactic extinction law with $R_V=3.1$ \citep{cardelli89}.
 Furthermore, estimates of $E(B-V)$ values of red quasars can be quite uncertain, where the difference in $E(B-V)$
 estimates can be as large as $\Delta E(B-V)=1.0-2.0$ mag for different methods \citep{glikman07,urrutia12}.
 The amount of extinction can also depend on dust properties such as grain sizes and the temperature of the dust,
 making it more difficult to make an accurate correction for it.
 Between color excess values of $E(B-V)=$ 1 and 2, the amounts of the extinction in H$\beta$ and H$\alpha$ line fluxes
 are different by factors of 30 and 10, respectively.

 However, the Paschen and Brackett hydrogen lines in the NIR and mid-infrared (MIR) mitigate the disadvantages of the UV/optical measurements.
 In the case of $E(B-V)=2$ mag, P$\beta$, P$\alpha$, Br$\beta$, and Br$\alpha$ line fluxes would be suppressed by factors of
 3.97, 2.16, 1.62, and 1.31, respectively, which are much less than those for the H$\beta$ and the H$\alpha$ lines.
 Consequently, the uncertainties of the NIR flux estimates due to $E(B-V)$ errors are significantly reduced.
 For these reasons, several NIR BH mass estimators are established by using Paschen \citep{kim10,landt13}
 and Brackett series lines \citep{kim15}.

 In this paper, we estimate $M_{\rm BH}$ and $L_{\rm bol}$ values for 16 red quasars at $z\sim 0.7$.
 The $M_{\rm BH}$ and the $L_{\rm bol}$ values are estimated with P$\beta$ (1.28\,$\mu$m) line properties \citep{kim10},
 and we compare the accretion rates of the red quasars to those of normal type 1 quasars,
 showing that accretion rates are much higher for red quasars.
 Throughout this paper, we use a standard $\Lambda$CDM cosmological model of $H_{0}$=70 km s$^{-1}$ Mpc$^{-1}$,
 $\Omega_{m}$=0.3, and $\Omega_{\Lambda}$=0.7 (e.g., \citealt{im97}).

\section{SAMPLE AND OBSERVATION}
 \subsection{The Sample}
 Our sample is drawn from red quasars listed in previous studies \citep{glikman07,urrutia09}.
 The previous studies selected red quasar candidates by using a combination of red colors in the optical through NIR
 (e.g., $R-K>4$ and $J-K>1.7$ in \citealt{glikman07}; $r^{\prime}-K>5$ and $J-K>1.3$ in \citealt{urrutia09})
 and FIRST \citep{becker95} radio detection. Then, they confirmed red quasars with spectroscopic follow-up.
 The previous studies found $\sim$80 red quasars, whose redshifts ranged from 0.186 to 3.050.
 These red quasars are very luminous (-31.61$<M_{i}<$-23.19; e.g., \citealt{urrutia09}) and obscured by dust (0.186$\leq E(B-V) \leq$ 3.05),
 and they show strong evidence of ongoing interaction \citep{urrutia08,glikman15}.
 Moreover, their spectra are well-fitted by normal type 1 quasar spectrum with a dust reddening law.

 \begin{figure}
 \centering
 \figurenum{1}
 \includegraphics[scale=0.4]{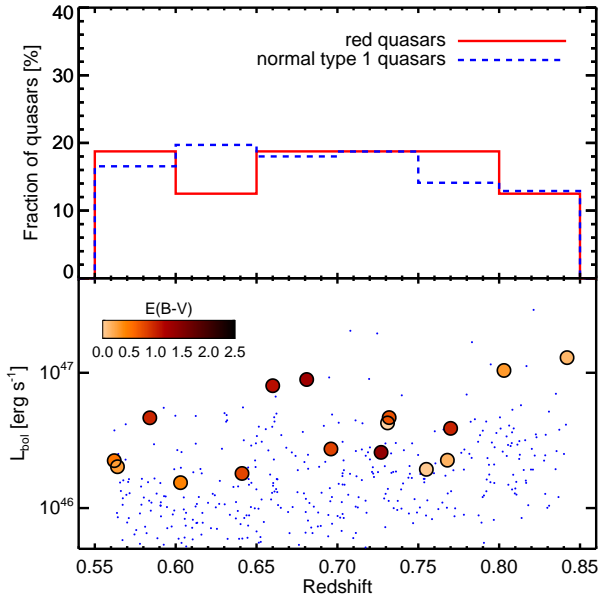}\\
 \caption{
 Top: redshift distributions of red quasars and normal type 1 quasars normalized by the sample size.
 The red solid and the blue dashed histograms represent the red quasars and normal type 1 quasars, respectively.
 Bottom: $L_{\rm bol}$ vs. redshift for red quasars and normal type 1 quasars.
 The circles and the blue dots represent red quasars and normal type 1 quasars, respectively,
 and the colors in the circles indicate their dust reddening values as shown in the legend.
 The $L_{\rm bol}$ values of the red quasars come from P$\beta$ luminosities,
 and L3000 and L5100 are used for the $L_{\rm bol}$ values of the normal type 1 quasars.
 }
 \end{figure}

 Among the red quasars, we select 20 red quasars at $z\sim0.7$ (from 0.56 to 0.84) for our study 
 where the redshifted P$\beta$ line is observable in the sky window at the $K$-band.
 Moreover, our sample is bright ($K<$15.5 mag) and has a wide range of luminosities (-31.78$<M_{K}<$-26.90 mag). 

 In order to compare the accretion rates of the red quasars to those of normal type 1 quasars,
 we select type 1 quasars from the quasar catalog \citep{schneider10} of
 the Sloan Digital Sky Survey (SDSS) Seventh Data Release (DR7; \citealt{abazajian09}).
 The catalog contains 105,783 spectroscopically confirmed quasars with wide ranges of redshifts from 0.065 to 5.4,
 and $r^{\prime}$ band magnitudes of 15.2$<r^{\prime}<$22.9.
 In order to avoid sample bias effects, we choose quasars with the same selection criteria as the red quasars except for the red colors:
 (i) the same redshift range of 0.56$\leq z \leq$0.84; (ii) FIRST radio detection; and (iii) detection in the $J$-, $H$-, and $K$-bands from 2MASS.
 In the end, 410 SDSS quasars are found and used as the control sample.
 Figure 1 shows the redshift distribution and the bolometric luminosity (see Section 4.2) vs. redshift of these 16 red quasars and 410 normal type 1 quasars.
 Note that the red quasars have a bolometric luminosity in the range of $\sim 10^{46}$--$10^{47}$\,$\rm erg~s^{-1}$.

 \begin{deluxetable}{cccccc}
 \tabletypesize{\scriptsize}
 \tablewidth{0pt}
 \tablenum{1}
 \tablecaption{Object List\label{tbl1}}
 \tablehead{
 \colhead{Objects}&	\colhead{Redshift}&	\colhead{$E(B-V)$}&	\colhead{$K$}&		\colhead{$r^{\prime}-K$}&	\colhead{$J-K$}\\
 \colhead{}&		\colhead{($z$)}&	\colhead{}&			\colhead{(mag)}&	\colhead{(mag)}&			\colhead{(mag)}}
 \startdata
0825$+$4716&	0.803&	0.52&	14.11&	6.42&	2.98\\
0911$+$0143&	0.603&	0.63&	14.87&	6.28&	1.59\\
0915$+$2418&	0.842&	0.36&	13.79&	6.47&	2.74\\
1113$+$1244&	0.681&	1.41&	13.67&	6.06&	2.47\\
1227$+$5053&	0.768&	0.38&	14.61&	4.16\tablenotemark{a}&	1.71\\
1248$+$0531&	0.740&	0.26&	14.63&	6.06&	3.45\\
1309$+$6042&	0.641&	0.95&	14.80&	6.77&	3.09\\
1313$+$1453&	0.584&	1.12&	14.48&	5.65&	2.30\\
1434$+$0935&	0.770&	1.14&	14.90&	6.19&	2.10\\
1532$+$2415&	0.562&	0.70&	14.96&	5.13\tablenotemark{a}&	1.87\\
1540$+$4923&	0.696&	0.97&	15.18&	5.31&	2.28\\
1600$+$3522&	0.707&	0.22&	14.17&	4.42\tablenotemark{a}&	2.24\\
1656$+$3821&	0.732&	0.88&	15.09&	6.91&	2.26\\
1720$+$6156&	0.727&	1.50&	15.20&	6.44&	1.95\\
2325$-$1052&	0.564&	0.50&	15.24&	4.46\tablenotemark{a}&	2.14\\
2339$-$0912&	0.660&	1.25&	14.09&	5.16&	1.63\\
 \enddata
\tablecomments{a: $R-K$ mags from \cite{glikman07}}
\end{deluxetable}

 \subsection{Observations}
 NIR spectra of the 20 red quasars were obtained
 using the SpeX instrument \citep{rayner03} on the NASA Infrared Telescope Facility (IRTF). 
 Descriptions of the observations of 11 of the 20 red quasars are given in \citet{glikman07,glikman12}.
 Additionally, we obtained IRTF NIR spectra for the remaining nine sources.
 In the observation, we used the short cross-dispersion mode (SXD: 0.8--2.5\,$\mu$m), whose spectral wavelength range includes redshifted P$\beta$ lines.
 We performed observation with a 0\farcs8 slit width to achieve a resolution of $R \sim$750 (400\,km s$^{-1}$ in FWHM),
 which is sufficient for the measurement of the width of broad emission line.
 Note that this observational setup is nearly identical to that of \citet{glikman07,glikman12}.

 For telluric correction, we obtained the spectrum of an A0V star after the observation of each target.
 The A0V stars were chosen to be near the red quasar ($\Delta{\rm airmass}< 0.1$ and separation $<15^{\circ}$).
 We use the Spextool \citep{vacca03,cushing04} package
 to produce fully reduced spectra.
%

 Our observations were performed under clear weather conditions and with a sub-arcsecond seeing of $\sim$0\farcs7,
 and we detect P$\beta$ emission line for five red quasars (0911$+$0143, 1248$+$0531, 1309$+$6042, 1313$+$1453, and 1434$+$0935)
 with a signal-to-noise ratio (S/N) of $>5$.
 We do not detect P$\beta$ line in the remaining four red quasars due to the low S/N of our data (1106$+$2812, 1159$+$2914, 1415$+$0903, and 1523$+$0030),
 and they are excluded in the following analysis.
 In total, we use 16 red quasars at $z \sim 0.7$ (from 0.56 to 0.84) among 20 quasars for which we got IRTF spectra.
 Table 1 summarizes the basic information of these 16 red quasars.

\section{ANALYSIS}
 We shift the fully reduced spectra to the rest frame,
 and correct the spectrum by adopting $E(B-V)$ values from previous studies \citep{glikman07,urrutia09}
 in the rest frame for each object. The $E(B-V)$ values were determined by comparing the spectra of the red quasars to
 the composite spectrum of quasars from the FIRST Bright Quasar Survey (\citealt{brotherton01})
 combined with a NIR quasar composite \citep{glikman06}.
 For the reddening law, these studies used the relation from \citet{fitzpatrick99}
 based on an average Galactic extinction curve from the IR through the UV
 with the conditions of $R_V=3.1$ and the elimination of a 2175\,$\rm \AA$ bump, which makes it similar to the Small Magellanic Cloud extinction curve.
 Although \citet{glikman07} provides the $E(B-V)$ values by using the Balmer decrement method for two targets, we use the $E(B-V)$ values
 by continuum fit because that the continuum based values are available for all of the objects.

 \begin{figure*}
 \centering
 \figurenum{2}
 \includegraphics[scale=0.4]{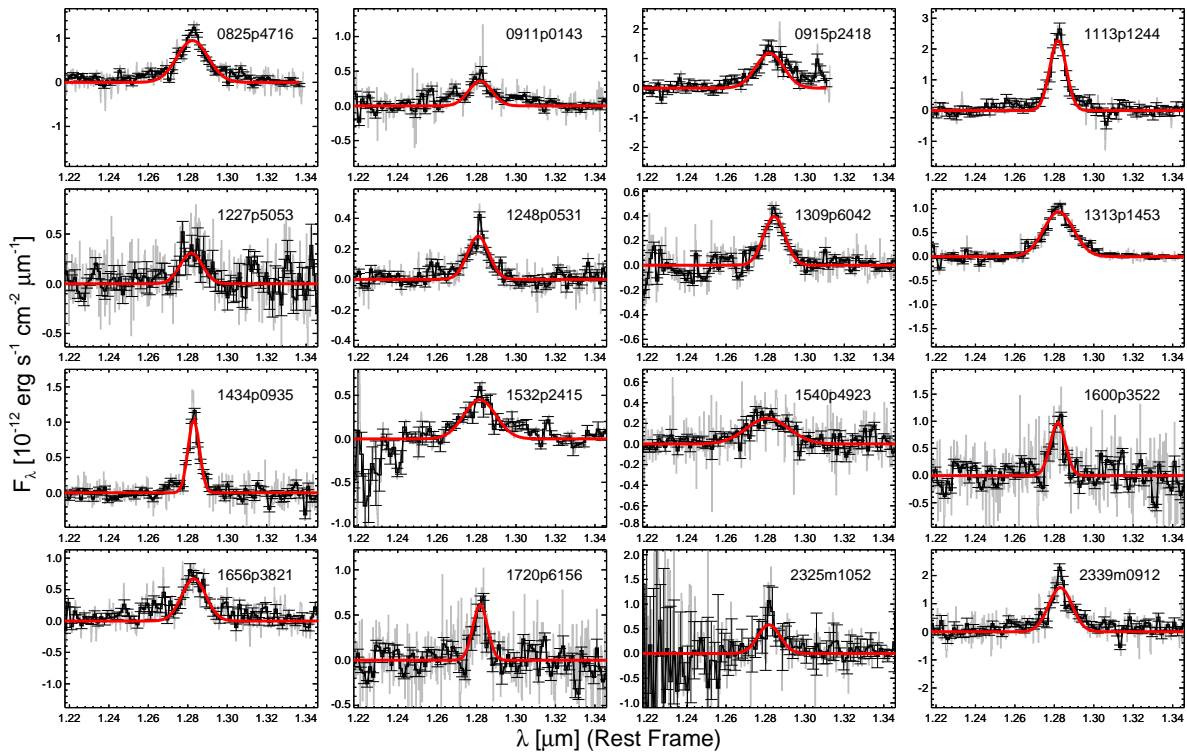}\\
 \caption{Results of the fitting of P$\beta$ lines with a Gaussian function.
 The continuum is already subtracted.
 The gray lines indicate observed spectra in the rest frame, and the black lines and the bars indicate binned spectra
 with a spectral resolution of $R\sim 750$ and associated errors.
 The red solid lines represent the best-fit model.
 }
 \end{figure*}

 After that, we determine the continuum around the P$\beta$ line
 to estimate the P$\beta$ line full width half maximum (FWHM) and luminosity.
 The continuum-fitting regions are chosen as 1.21--1.25\,$\mu$m and 1.31--1.35\,$\mu$m, with a width of 0.4\,$\mu$m.
 The continuum-fitting regions are 0.03\,$\mu$m away from the P$\beta$ line,
 where the width of 0.03\,$\mu$m corresponds to $\sim$7000\,km\,s$^{-1}$, which is similar to 2$\times$FWHM of a broad-line of a typical type 1 quasar.
 We fit the continuum with a linear function because the NIR spectrum around P$\beta$ line can be approximated with a power-law component
 over a limited wavelength range \citep{glikman06,kim10,kim15}.

 After the continuum subtraction, we fit the P$\beta$ emission line with a single Gaussian function.
 An IDL procedure, \texttt{MPFITEXPR} \citep{markwardt09} code was used to fit the line.
 For the fit, we set free the central wavelength, integrated flux, and FWHM of the Gaussian component.
 The fit provides the FWHM and the integrated flux of the P$\beta$ line,
 and the measured FWHM values were corrected for
 the dispersion in the wavelength due to the spectral resolution of the instrument through an equation of
 ${\rm FHWM = \sqrt{{FWHM_{obs}}^2 - {FWHM_{inst}}^2}}$ where ${\rm FWHM_{inst} = 400\,km\,s^{-1}}$.
 The single component fitting can introduce a small systemic bias in the derived values of FWHM and line luminosities.
 To correct for this bias, we adopted the correction factors of ${\rm FWHM}_{\rm multi}$/${\rm FWHM}_{\rm single}$=0.91 and
 ${L}_{\rm multi}$/${L}_{\rm single}$=1.06, which are derived from higher-resolution Balmer line profiles of
 other normal type 1 active galactic nuclei (AGNs; \citealt{kim10}; see also the discussion in \citealt{kim15}).

 The formal fitting uncertainties are found to be 4.7\,\% (from 1.0\,\% to 24.6\,\%) and 6.9\,\% (from 1.3\,\% to 34.7\,\%)
 for the FWHM and the flux, respectively. Several high ($>10$\,\%) formal fitting uncertainties are due to the fact that
 the observed P$\beta$ line is located at the edge of the spectral window of SpeX (for 0915$+$2418),
 or the spectrum has a low S/N (for 1227$+$5053 and 2325$-$1052).
 An additional uncertainty is considered since the values of the fitted parameter vary with how we define the continuum level.
 We varied the wavelength regions in the vicinity of the P$\beta$ line for the continuum fit, and
 we find that the variation of the continuum level yields to a change of 6.0\,\% (from 0.9\,\% to 36.9\,\%) in the FWHM values
 and 6.9\,\% (from 0.9\,\% to 43.9\,\%) in the flux values.
 Overall, we combined these two uncertainties in the quadrature ($\sigma_{\rm tot}^{2}=\sigma_{\rm a}^{2}+\sigma_{\rm b}^{2}$).
 The combined uncertainties of the FWHM and the flux are 10.0\,\% (from 2.4\,\% to 38.0\,\%) and 10.6\,\% (from 3.1\,\% to 45.5\,\%), respectively,
 which corresponds to 0.10 dex in $M_{\rm BH}$ value.

\section{BH Masses and Bolometric Luminosities}
 \subsection{BH Masses}
 For estimating virial masses of red quasars, we adopted the P$\beta$ scaling relation \citep{kim10}
 after adjusting the relation to have a recent virial coefficient of $\log f = 0.05$ \citep{woo15}.
 The modified relation is
 \begin{equation}
\frac{M_{\rm BH}}{M_{\odot}}=10^{7.04\pm0.02} \left( \frac{L_{\rm P\beta}}{\rm 10^{42}\,erg\,s^{-1}} \right)^{0.48\pm0.03}
\left( \frac{\rm FWHM_{\rm P\beta}}{\rm 10^{3}\,km\,s^{-1}} \right)^{2}. 
 \end{equation}
 In this scaling relation, the luminosity and the FWHM of the P$\beta$ line represent the radius and the velocity
 of the BLR.
 The measured $M_{\rm BH}$ values of the red quasars are listed in Table 2.

 For the BH masses of normal type 1 quasars, we use optical or UV spectral properties from \cite{shen11}
 and use the $M_{\rm BH}$ estimators of \cite{mclure04}.
 Note that we modified their formula so that it gives $\log f=0.05$ by multiplying by a factor of 1.122.
 First, we use $\lambda L_{\rm 5100\AA}$ (L5100, hereafter) and FWHM of the H$\beta$ line with the equation of 
 \begin{equation}
\frac{M_{\rm BH}}{M_{\odot}}=5.27 \left( \frac{{\rm L5100}}{\rm 10^{44}\,erg\,s^{-1}} \right)^{0.61}
\left( \frac{\rm FWHM_{\rm H\beta}}{\rm km\,s^{-1}} \right)^{2}. 
 \end{equation}
 We use this $M_{\rm BH}$ estimator for 406 normal type 1 quasars with the values of L5100 and $\rm FWHM_{H\beta}$ from \cite{shen11}.
 For the remaining four normal type 1 quasars with no H$\beta$ detection, we adopt \ion{Mg}{2} and $\lambda L_{\rm 3000\AA}$ (L3000) instead as
 \begin{equation}
\frac{M_{\rm BH}}{M_{\odot}}=3.59 \left( \frac{{\rm L3000}}{\rm 10^{44}\,erg\,s^{-1}} \right)^{0.62}
\left( \frac{\rm FWHM_{\rm MgII}}{\rm km\,s^{-1}} \right)^{2}. 
 \end{equation}

 \subsection{Bolometric Luminosities}
 To obtain bolometric luminosities of red quasars, we use several relations between
 the line luminosity, the continuum luminosity, and the bolometric luminosity.
 We bootstrap the relations between $L_{\rm P\beta}$ and $L_{\rm H\alpha}$ \citep{kim10},
 $L_{\rm H\alpha}$ and L5100 \citep{jun15}, and L5100 and $L_{\rm bol}$ \citep{shen11} to relate $L_{\rm P\beta}$ to $L_{\rm bol}$.
 The combined relationship is 
 \begin{equation}
 \log(\frac{L_{\rm bol}}{\rm{10^{44}\,erg\,s^{-1}}})=1.29+0.969\,\log(\frac{L_{\rm P\beta}}{\rm{10^{42}\,erg\,s^{-1}}}).
 \end{equation}
 The measured $L_{\rm bol}$ values of the red quasars are listed in Table 2.

 For the normal type 1 quasars for which BH masses are estimated by a H$\beta$ BH mass estimator,
 we convert L5100 to the bolometric luminosity with a bolometric correction factor of 9.26 \citep{shen11} for L5100.
 We also measure the bolometric luminosities of four normal type 1 quasars without H$\beta$ detection from
 L3000 with a bolometric correction factor of 5.15 \citep{shen11}.

 \begin{deluxetable*}{cccccccc}
 \tabletypesize{\scriptsize}
 \tablewidth{0pt}
 \tablenum{2}
 \tablecaption{P$\beta$ parameters, BH masses, and Eddington ratios\label{tbl2}}
 \tablehead{
 \colhead{Objects}&
 \colhead{$\rm FWHM_{\rm P\beta}$}& \colhead{$\log (L_{\rm P\beta}$)}& \colhead{$M_{\rm BH}$}&						\colhead{$L_{\rm bol}$}&
 \colhead{Eddington Ratio}& \colhead{$\dot{M}$}&							\colhead{$\tau$}\\
 \colhead{}&
 \colhead{($\rm km\,s^{-1}$)}&			\colhead{($\rm erg\,s^{-1}$)}& \colhead{(${\rm 10^{8}}$\,$M_{\rm\odot}$)}&	\colhead{(10$^{46}$\,erg\,s$^{-1}$)}&
 \colhead{}&				\colhead{($M_{\rm\odot}$\,$\rm year^{-1}$)}&	\colhead{($10^{7}\,$years)}}
 \startdata
0825$+$4716&	3940$\pm$393&	43.78$\pm$0.04&	12.25$\pm$2.98&	10.40$\pm$1.02&	0.693&	18.45&	6.64\\ 
0911$+$0143&	2907$\pm$206&	42.93$\pm$0.04&	2.59$\pm$0.44&	1.54$\pm$0.16&	0.486&	2.73&	9.48\\ 
0915$+$2418&	3527$\pm$729&	43.88$\pm$0.09&	10.94$\pm$4.92&	12.95$\pm$3.18&	0.967&	22.96&	4.76\\ 
1113$+$1244&	2054$\pm$107&	43.71$\pm$0.03&	3.09$\pm$0.52&	8.92$\pm$0.67&	2.360&	15.82&	1.95\\ 
1227$+$5053&	2781$\pm$446&	43.10$\pm$0.08&	2.87$\pm$1.00&	2.26$\pm$0.49&	0.644&	4.01&	7.16\\ 
1248$+$0531&	2697$\pm$123&	43.03$\pm$0.02&	2.49$\pm$0.32&	1.93$\pm$0.12&	0.633&	3.43&	7.27\\ 
1309$+$6042&	2673$\pm$89&	43.00$\pm$0.02&	2.37$\pm$0.26&	1.80$\pm$0.09&	0.622&	3.20&	7.40\\ 
1313$+$1453&	3793$\pm$96&	43.42$\pm$0.01&	7.62$\pm$0.92&	4.65$\pm$0.17&	0.499&	8.25&	9.24\\ 
1434$+$0935&	1398$\pm$34&	43.34$\pm$0.01&	0.95$\pm$0.11&	3.88$\pm$0.13&	3.350&	6.89&	1.37\\ 
1532$+$2415&	4071$\pm$1549&	43.09$\pm$0.16&	6.12$\pm$4.87&	2.24$\pm$1.05&	0.299&	3.98&	15.39\\ 
1540$+$4923&	5397$\pm$1518&	43.18$\pm$0.14&	11.86$\pm$7.13&	2.73$\pm$1.11&	0.188&	4.85&	24.46\\ 
1600$+$3522&	1887$\pm$228&	43.38$\pm$0.06&	1.81$\pm$0.49&	4.26$\pm$0.63&	1.927&	7.55&	2.39\\ 
1656$+$3821&	3028$\pm$246&	43.42$\pm$0.04&	4.87$\pm$0.98&	4.66$\pm$0.53&	0.783&	8.27&	5.88\\ 
1720$+$6156&	1818$\pm$205&	43.16$\pm$0.06&	1.31$\pm$0.33&	2.58$\pm$0.38&	1.613&	4.58&	2.86\\ 
2325$-$1052&	2778$\pm$772&	43.05$\pm$0.13&	2.71$\pm$1.59&	2.02$\pm$0.73&	0.610&	3.58&	7.55\\ 
2339$-$0912&	2927$\pm$140&	43.67$\pm$0.03&	5.95$\pm$0.95&	8.03$\pm$0.55&	1.102&	14.23&	4.18\\ 
 \enddata
 \end{deluxetable*}

\subsection{Comparison of the $L_{\rm bol}$ and the $M_{\rm BH}$ from Different Estimators}
 Five of the red quasars in our sample overlap with the thirteen red quasars studied by \citet{urrutia12}.
 We are thus able to compare the $L_{\rm bol}$ and the $M_{\rm BH}$ values of
 these five red quasars (0825$+$4716, 0915$+$2418, 1113$+$1244, 1532$+$2415, and 1656$+$3821)
 from our method versus the method adopted by \cite{urrutia12}.
 \citet{urrutia12} obtained $L_{\rm bol}$ by applying a bolometric correction to their
 monochromatic luminosity at 15\,$\mu$m (L15, hereafter),
 and derived $M_{\rm BH}$ values using L15 as a proxy for $R_{\rm BLR}$
 and FWHM of broad emission lines (either Balmer or Paschen lines depending on the availability).
 For comparison, the $M_{\rm BH}$ values of \citet{urrutia12} are rescaled to match the virial factor adopted in this paper.
 Figure 3 shows a comparison of $L_{\rm bol}$ and $M_{\rm BH}$, which show a reasonable agreement between two independent estimates,
 except for one outlier (1532$+$2415, a red circle in Figure 3).
 The Pearson correlation coefficients are 0.863 and 0.370 for the $L_{\rm bol}$ and the $M_{\rm BH}$, respectively,
 but the Pearson correlation coefficient for $M_{\rm BH}$ values increases to 0.683 if we exclude the outlier.
 We note that the outlier has the largest uncertainty of the P$\beta$ line properties among the 16 red quasars used in this study,
 with the fractional errors of the FWHM and the flux as 38\,\% and 45.5\,\%, respectively.

 \begin{figure*}
 \centering
 \figurenum{3}
 \includegraphics[scale=0.5]{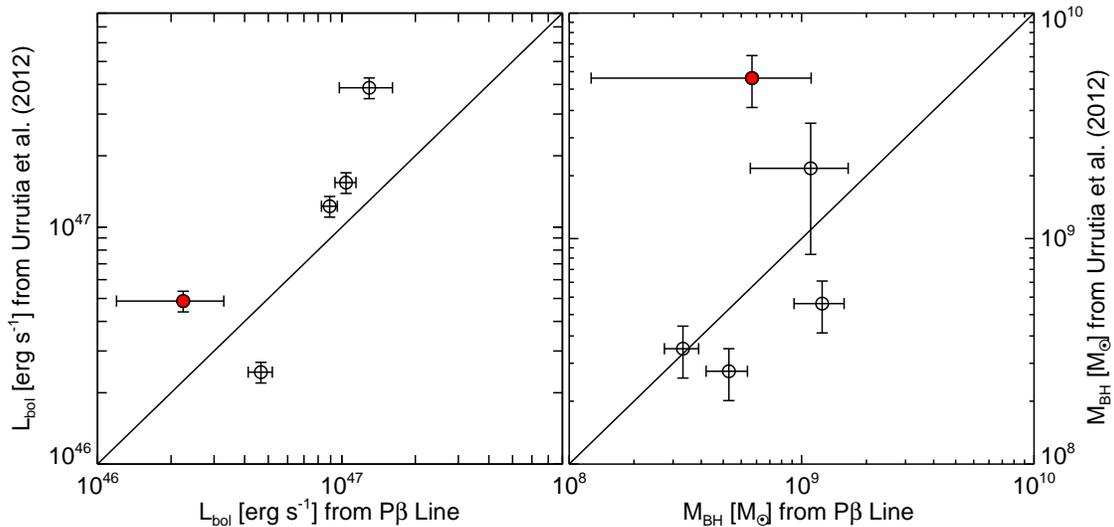}\\
 \caption{
 (Left): comparison of $L_{\rm bol}$ values derived from P$\beta$ (this work) vs. $L_{\rm bol}$ values derived from L15 \citep{urrutia12}.
 The black solid line indicates the case where the $L_{\rm bol}$ values from the two methods are identical.
 The outlier (1532$+$2415, red circle) has the largest uncertainties of the FWHM and the luminosity among P$\beta$-based measurements.
 (Right): comparison of $M_{\rm BH}$ values derived from P$\beta$ (this work) vs.
 $M_{\rm BH}$ values derived from L15 and FWHM of broad emission lines \citep{urrutia12}.
 The meaning of the lines and the red circle are identical to the left panel.
 The $L_{\rm bol}$ and $M_{\rm BH}$ values agree to within 0.31 dex and 0.48 dex, respectively.
 }
 \end{figure*}

 The rms scatters (including measurement uncertainties) of the $L_{\rm bol}$ and the $M_{\rm BH}$ values
 with respect to the one-to-one correlations are 0.306 dex and 0.479 dex, respectively.
 In order to derive the intrinsic scatters by removing the contribution from measurement uncertainties,
 we performed a Monte-Carlo simulation 10000 times by assuming that the $L_{\rm bol}$ and the $M_{\rm BH}$ values
 from \citet{urrutia12} are identical to those from P$\beta$ measurements
 and adding measurement uncertainties randomly.
 The median values of the scatters from this simulation are taken to be the contribution to the rms scatter due to
 measurement uncertainties, and they are subtracted from the original rms scatters in quadrature to obtain intrinsic scatter.
 Through this process, we find that the intrinsic scatters in the correlations
 are 0.294 dex and 0.446 dex, respectively, suggesting that the scatters in the two quantities
 are not due to measurement errors.

 In Figure 4, we compare the Eddington ratios from P$\beta$ to those from \citet{urrutia12}.
 The comparison shows a reasonable agreement between two values with an rms scatter of 0.514 with respect to the one-to-one correlation.
 However, if the outlier is excluded, the Eddington ratios from \citet{urrutia12} are
 a bit larger than those from P$\beta$ by a factor of 1.85 (0.267 dex).

 \begin{figure}
 \centering
 \figurenum{4}
 \includegraphics[scale=0.5]{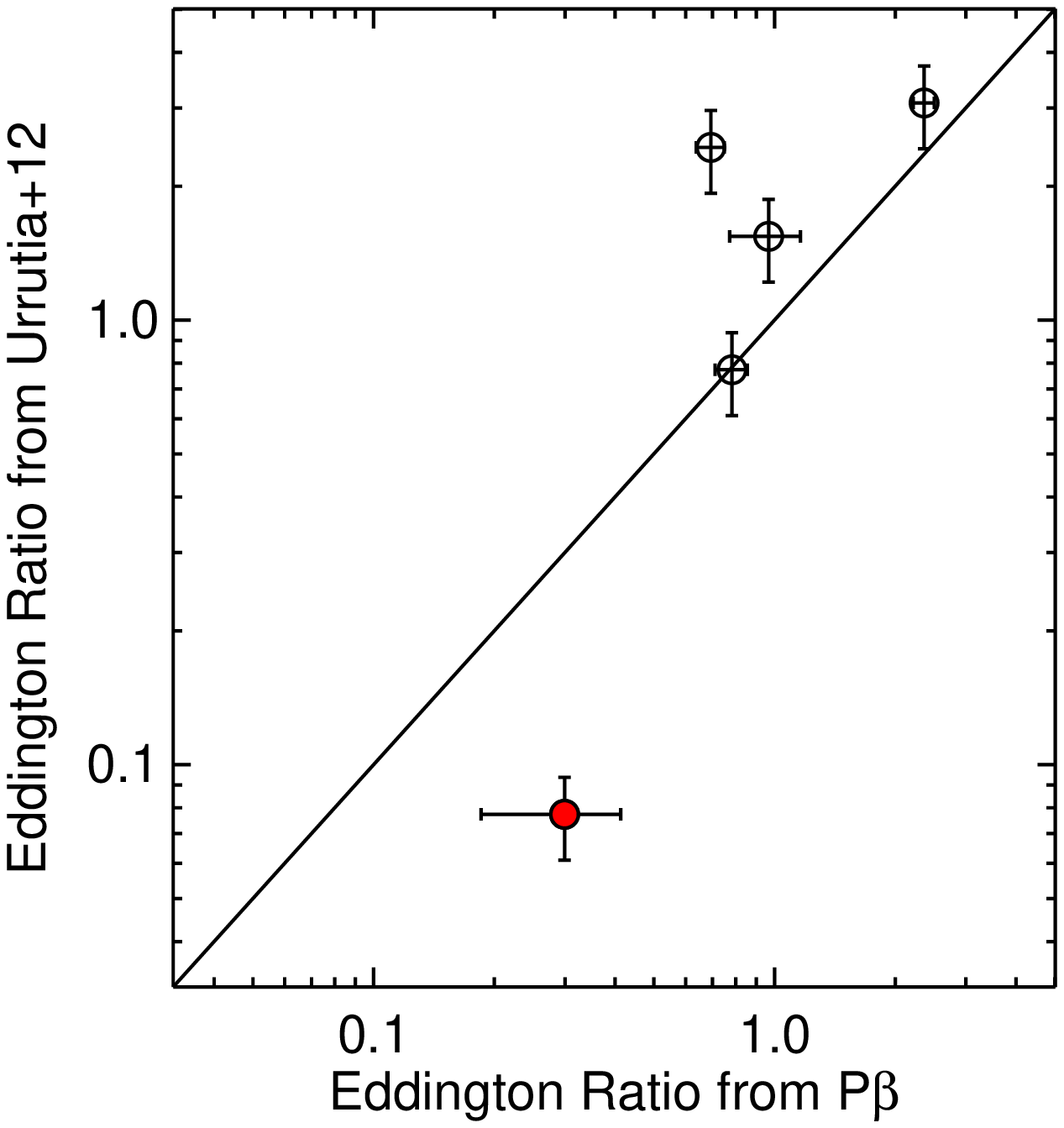}\\
 \caption{Comparison of the Eddington ratios from \citet{urrutia12} and those from P$\beta$.
 The black solid line indicates the case where the Eddington ratios are identical, and
 the meaning of the red circle is identical to Figure 3.
 If the outlier is excluded, the Eddington ratios from \cite{urrutia12} are a bit larger than those from P$\beta$.
 }
 \end{figure}

\section{DISCUSSION}
\subsection{Accretion Rates of Red Quasars}
 In order to characterize the accretion rates of red quasars, we compare the distribution of the Eddington ratios
 ($L_{\rm bol}$/$L_{\rm Edd}$, where $L_{\rm Edd}$ denotes the Eddington luminosity)
 of red quasars and those of normal type 1 quasars. The median Eddington ratios of the red quasars and the normal type 1 quasars are
 0.69 and 0.16, respectively. 
 Moreover, Figure 6 shows that the distributions of the Eddington ratios of the red quasars and the normal type 1 quasars are clearly different,
 with the red quasars having larger $L_{\rm bol}$/$L_{\rm Edd}$ values than the normal type 1 quasars.
 In order to quantify how significantly these two distributions differ from each other,
 we perform the Kolmogorov-Smirnov test (K--S test).
 For the K--S statistics, the maximum deviation between the cumulative distributions
 of these two Eddington ratios, $D$, is 0.71, and the probability of the null hypothesis in which
 $L_{\rm bol}$/$L_{\rm Edd}$ values of the two samples are drawn from the same distribution,
 $p$, is only $9.07 \times 10^{-8}$, confirming that the Eddington ratios of the red quasars
 are clearly skewed toward higher values in comparison to those of the normal type 1 quasars.

 \begin{figure}
 \centering
 \figurenum{5}
 \includegraphics[scale=0.45]{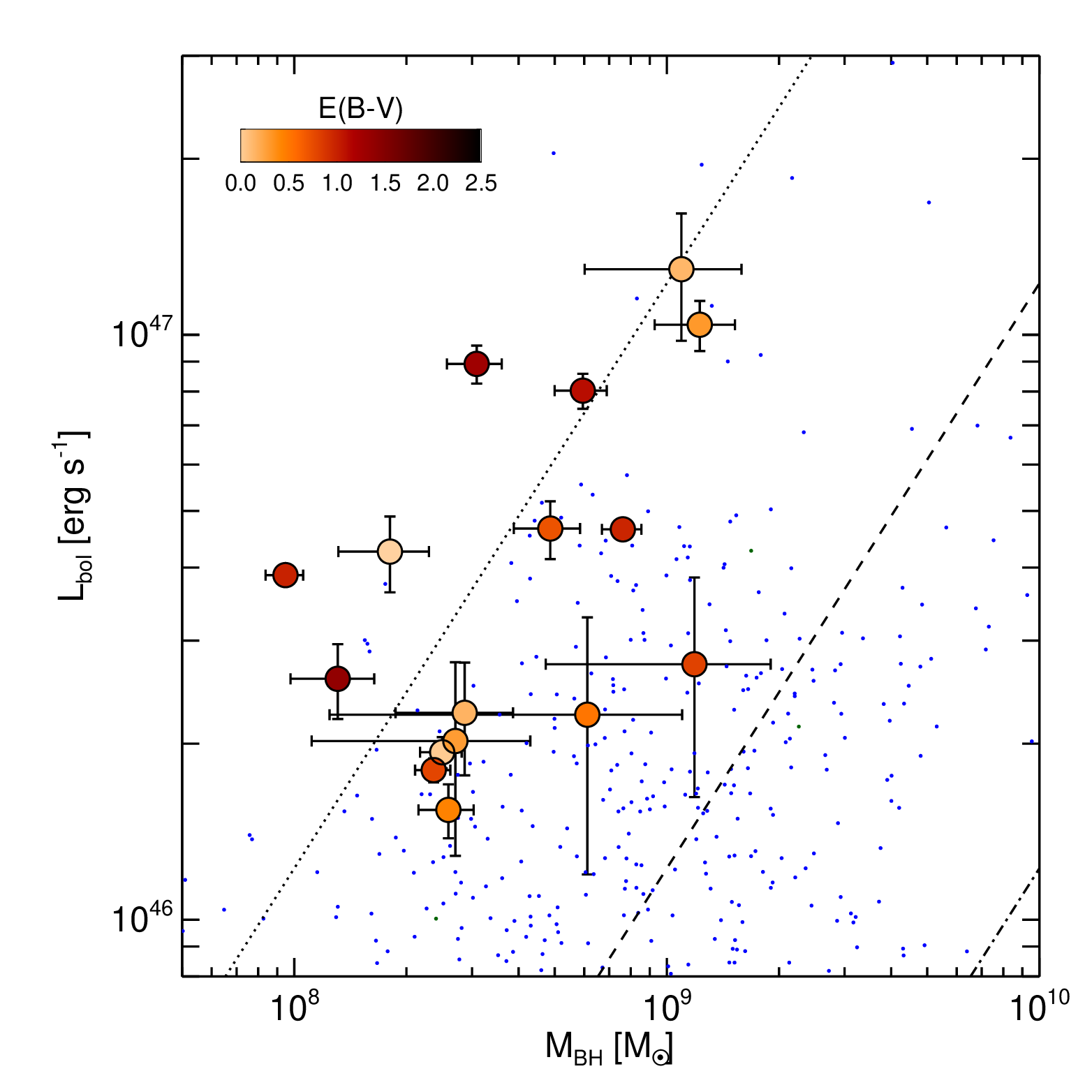}\\
 \caption{The $L_{\rm bol}$ versus $M_{\rm BH}$ of red quasars and normal type 1 quasars.
 The red quasars are shown with open circles, and the colors in the circles represent their dust reddening values.
 The blue dots indicate the $L_{\rm bol}$ and the $M_{\rm BH}$ values of normal type 1 quasars.
 The dotted, dashed, and dash-dotted lines indicate Eddington ratios of 1.0, 0.1, and 0.01, respectively.
 }
 \end{figure}

 Because Eddington ratios can vary with BH mass,
 we also show the Eddington ratio distribution of the sample divided into BH mass bins
 to account for any difference in the Eddington ratio distribution arising from the BH mass effect (e.g., \citealt{shen11}).
 We divided the red quasars and the normal type 1 quasars into a low-mass ($7.9<\log(M_{\rm BH}/M_{\odot})<8.5$)
 and a high-mass ($8.5<\log(M_{\rm BH}/M_{\odot})<9.1$) sample.
 For the low-mass sample, 9 red quasars and 67 normal type 1 quasars are selected, and
 their $\log(L_{\rm bol}/{\rm erg\,s^{-1}})$ values have ranges of 46.19--46.95 and 45.36--46.57
 for the red quasars and the normal type 1 quasars, respectively.
 The $D$ and the $p$ values from a K--S test for these samples are 0.64 and 0.0015, respectively.
 The high-mass sample includes 7 red quasars and 177 normal type 1 quasars. The $\log(L_{\rm bol}/{\rm erg\,s^{-1}})$ values
 have ranges of 46.35--47.11 and 45.44--47.31 for the red quasars and the normal type 1 quasars, respectively.
 The $D$ value is 0.61, and the $p$ is 0.0074 between these two distributions.
 Again, we find that the Eddington ratio distribution of red quasars is significantly different from that of normal type 1 quasars,
 independent of the BH mass.
 The Eddington ratio values of red quasars are about 0.6--0.8 dex larger than those of normal type 1 quasars.

 \begin{figure}
 \centering
 \figurenum{6}
 \includegraphics[scale=0.45]{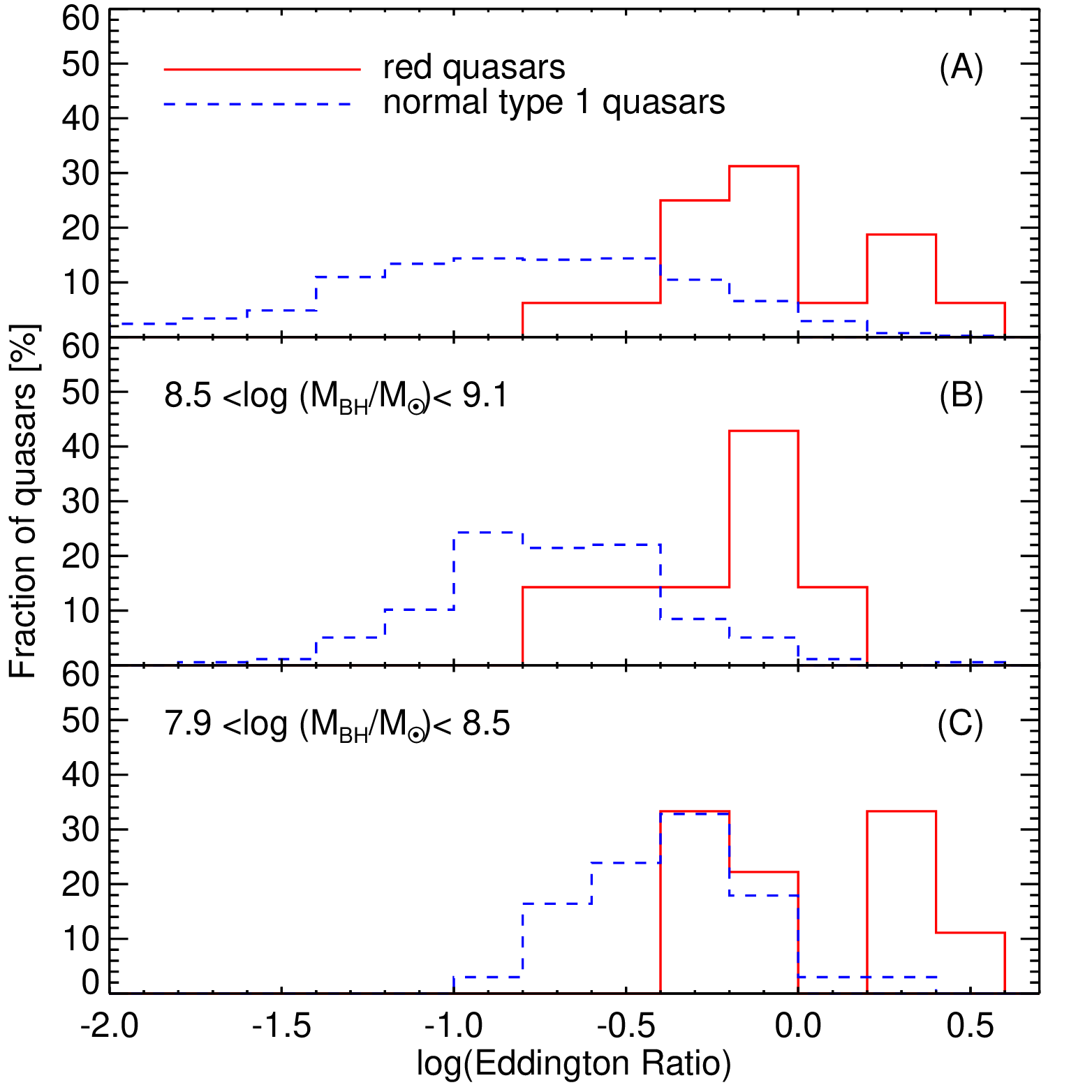}\\
 \caption{(A): distributions of the Eddington ratios of red quasars and normal type 1 quasars. The red solid and the blue dashed histograms
 represent the red quasars and normal type 1 quasars, respectively.
 (B): distributions of the Eddington ratios for the high BH mass sample ($10^{8.5} < {M_{\rm BH}}/{M_{\rm \odot}} < 10^{9.1}$).
 (C): distributions of the Eddington ratios for the low BH mass sample ($10^{7.9} < {M_{\rm BH}}/{M_{\rm \odot}} < 10^{8.5}$).
 }
 \end{figure}

 There are three potential biases that could affect the result--(i) the uncertainty in $E(B-V)$,
 (ii) the application of the $M_{\rm BH}$ estimator that is derived from normal type 1 quasars, and
 (iii) the omission of four red quasars without P$\beta$ detection in the analysis.
 We discuss below how much each item could affect the result.

 \begin{figure*}
 \centering
 \figurenum{7}
 \includegraphics[width=\textwidth]{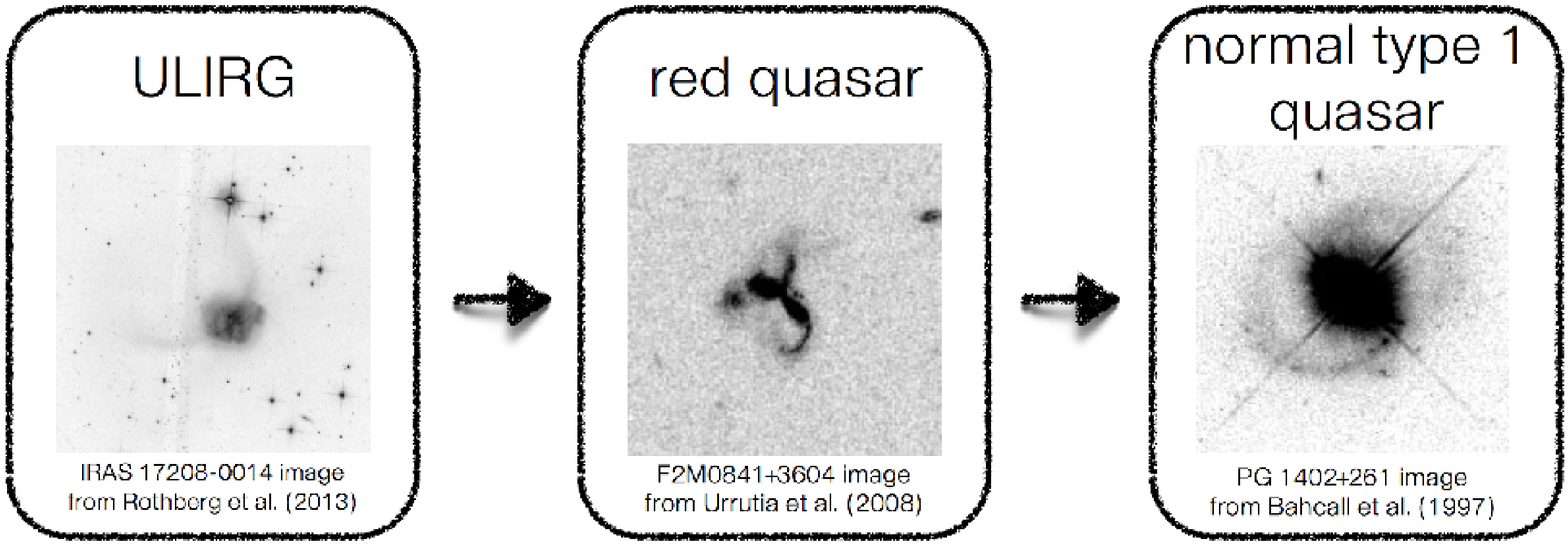}\\
 \caption{Schematic outline of the galaxy evolution scenario from ULIRG to normal type 1 quasars.
 Image credits belong to \citet{rothberg13}, \citet{urrutia08}, and \citet{bahcall97}.
 }
 \end{figure*}

 First, the Eddington ratios of the red quasars could be overestimated or underestimated,
 since the $E(B-V)$ values \citep{glikman07,urrutia12} of the red quasars can vary by as much as $\sim$ 0.5,
 depending on how they are derived.
 For example, for the case of 1227$+$5053 of which $E(B-V)$ value is 0.38 (continuum-derived) or -0.23 (Balmer decrement-derived)\footnotemark, 
 the extinction-corrected P$\beta$ luminosity can be over/underestimated by a factor of 1.6 depending on the method for estimating the $E(B-V)$ value.
 \footnotetext{\textbf{Although the $E(B-V)$ value based on the Balmer decrement, -0.23, is unphysical, the negative $E(B-V)$ value
 is due to the line flux measurement uncertainty, related to the low S/N of the spectrum.
 }}
 If we adopt an $E(B-V)$ value of -0.23 from the Balmer decrement, the Eddington ratio of 1227$+$5053 would decrease from 2.87 to 2.26.
 Note that this is a rather extreme case. Even if we decrease the Eddington ratios of all of 16 red quasars by a factor of 1.6,
 the median Eddington ratios of red quasars is 0.55, which is still significantly higher than that of normal type 1 quasars of 0.16.
 Therefore, the uncertainty in the reddening correction does not affect the result about
 the Eddington ratios of red quasars, and this demonstrates the power of using the NIR emission line in dusty systems,
 as outlined also in the introduction.

 Second, the P$\beta$ $M_{\rm BH}$ estimator that is derived from normal type 1 quasars
 may not be applicable to red quasars if the BLR physics of red quasars is very different from that of normal type 1 quasars
 to the extent that the virial $M_{\rm BH}$ estimator does not hold for red quasars.
 This is an open question, but we note that our red quasars are moderately obscured quasars for which the BLR is expected to be well-established.
 Therefore, the characteristics of the BLR physics of red quasars should be very similar to those of normal type 1 quasars.

 Third, we omitted four red quasars from our analysis (1106$+$2812, 1159$+$2914, 1415$+$0903, and 1523$+$0030)
 because their P$\beta$ lines were not detected with S/N $>$5, and this may bias the result.
 Among the four red quasars, we estimated L5100 and $\rm FWHM_{H\beta}$ values
 for 1106$+$2812, 1415$+$0903, and 1523$+$0030 using the optical spectra from \cite{glikman06,glikman12} and NIR spectra from our IRTF data,
 but we could not estimate the $\rm FWHM_{H\beta}$ for 1159$+$2914 due to low-quality data (S/N $<$3).
 From L5100 and $\rm FWHM_{H\beta}$, we computed $L_{\rm bol}$ and $M_{\rm BH}$ values and the Eddington ratios,
 and we found that the Eddington ratios of 1106$+$2812, 1415$+$0903, and 1523$+$0030 are 0.477, 0.186, and 0.266, respectively.
 If the Eddington ratios of the three red quasars are included in the Eddington ratio comparison,
 the $D$ and $p$ values from the K--S test are 0.63 and 4.34$\times 10^{-7}$, respectively,
 which are nearly identical to the result without the three red quasars.

 We conclude that red quasars have high accretion rates, with $L_{\rm bol}$/$L_{\rm Edd} \simeq 0.69$.
 This Eddington ratio is higher than normal type 1 quasars by a factor of 4--5.
 These results, although limited by the modest size of the sample used in this study, are consistent with
 the scenario that red quasars are in the intermediate stage between merger-driven star-forming galaxies and the normal type 1 quasars,
 rather than a scenario in which the red colors of red quasars originate from a viewing angle difference in an AGN unification scheme.
 Under the merger-driven quasar evolutionary scenario as depicted in Figure 7,
 the early phase of major mergers drives gas toward the central region of the galaxy
 due to the loss of the gas' angular momentum through a dissipative, violent merging.
 This process can fuel the BH activities and a rapid BH growth inside a dust-enshrouded host galaxy \citep{hopkins05,hopkins08}.
 Such a merger-driven quasar evolutionary scenario also has observational support, especially for luminous quasars (e.g., \citealt{hong15}).


\subsection{Duration of Red Quasar Phase}
 The duration period of quasar activity is uncertain, with a wide range of possible values of $10^{6}$--$10^{8}$ years \citep{martini04}.
 Quasars shine at the bolometric luminosity of $L_{\rm bol} =\epsilon \dot{M} c^{2}$,
 where $\epsilon$ is the radiation efficiency that has a value of $\epsilon \simeq 0.1$ for luminous quasars \citep{martini04}.
 Under the Eddington-limited accretion (maximal accretion),
 BH mass can grow exponentially as $\sim \exp(t/\tau_{S})$, where $\tau_{s}$ is
 the Salpeter time, $\tau_{S} = 4.5 \times 10^{7} (\epsilon/0.1)$ years.
 If the quasar lifetime is $< 10^{7}$ years, this is less than Salpeter time,
 and it cannot account for the growth of
 SMBHs to $M_{\rm BH} \sim 10^{9} M_{\rm \odot}$ from a seed BH with $M_{\rm BH} < 10^{6} M_{\rm \odot}$.
 Hence, it has been suggested that SMBHs underwent
 an extended growth period, including a red quasar phase
 that can last more than a few $\times 10^{8}$ years \citep{kelly10}.

 \begin{figure}
 \centering
 \figurenum{8}
 \includegraphics[scale=0.45]{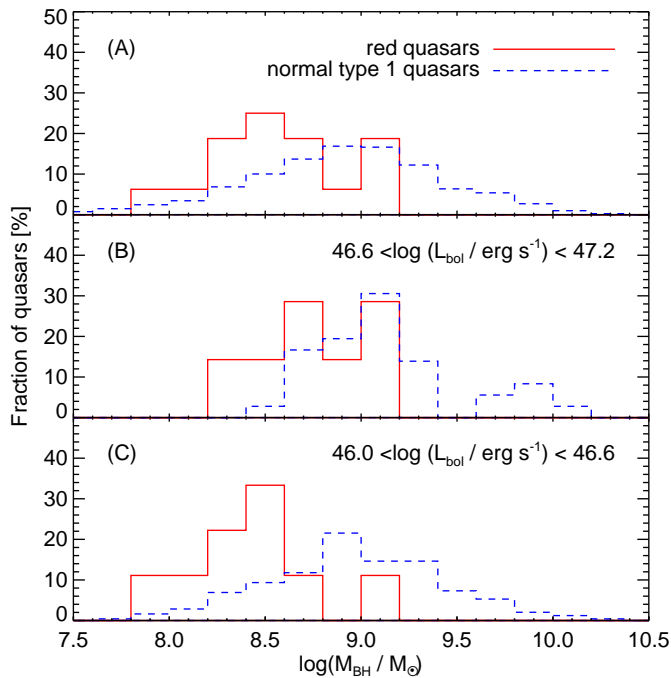}\\
 \caption{(A): $M_{\rm BH}$ distributions of red quasars and normal type 1 quasars. The red solid and the blue dashed histograms
 represent red quasars and normal type 1 quasars, respectively.
 (B): $M_{\rm BH}$ distributions for high-luminosity quasars ($10^{46.6} < {L_{\rm bol}}/{\rm erg~s^{-1}} < 10^{47.2}$).
 (C): $M_{\rm BH}$ distributions for low-luminosity quasars ($10^{46.0} < {L_{\rm bol}}/{\rm erg~s^{-1}} < 10^{46.6}$).
 }
 \end{figure}

 We can see if such a growth makes sense by estimating the red quasar duration time from the accretion rate and $M_{\rm BH}$ of our red quasar sample.
 Since the Eddington ratios are roughly 1 for red quasars, we can propose that the BH mass grows exponentially as
 $\sim \exp(t/\tau)$ and use the characteristic time $\tau$ as a rough estimate of
 the duration of the red quasar phase.
 Here, $\dot{M}$ is derived using the relations of $L_{\rm bol} =\epsilon \dot{M} c^{2}$ (assuming $\epsilon=0.1$),
 and $\tau = M_{\rm BH}$/$\dot{M}$.

 The derived $\dot{M}$ and $\tau$ values are listed in Table 2, 
 and we find that $\tau$ values range from $10^{7}$ to $2\times 10^{8}$ years,
 or $\sim 7 \times 10^{7}$ years on average.
 We see from Figure 8 that the average BH mass of normal type 1 quasars is a few times larger than that of red quasars.
 To grow the BH mass this much during the red quasar phase, we need only one to two
 Salpeter time, which supports the use of $\tau$ values as estimates of the duration of the red quasar phase.


 Our estimate of the red quasar duration time is $\sim$10 times larger than the value presented in \cite{glikman12}.
 The estimate in \cite{glikman12} relies on the fraction of red quasars to normal type 1 quasars ($\sim 20\%$) 
 and assumes a duration time of $10^{7}$ years for normal type 1 quasars.
 The difference can be reconciled if we adopt a duration time of $\sim 10^{8}$ years 
 for normal type 1 quasars as found in \cite{kelly10} and \cite{cao10}.
 Since the Eddington ratios of normal type 1 quasars are $\times 5$ times lower than those of 
 red quasars, the growth rate of BH masses in normal type 1 quasars is much lower than red quasars
 and the long duration time does not lead to the over-growth of BH mass.
 Alternatively, we can reconcile the two results if $\epsilon \ll 0.1$ during the red quasar phase.

\section{SUMMARY}
 We estimated the Eddington ratios of 16 red quasars at $z \sim 0.7$ ($0.56 < z < 0.84$)
 using the P$\beta$ line to minimize the extinction due to dust in red quasars.
 We compared the distribution of the Eddington ratios of the red quasars to those of normal type 1 quasars.
 For the Eddington ratios of normal quasars, 410 normal type 1 quasars at $z \sim 0.7$ in the SDSS DR7 quasar catalog were used.
 We show that the Eddington ratios of red quasars are significantly higher than those of normal type 1 quasars
 by a factor of 4.3.
 Furthermore, for a give Eddington ratio, the average $M_{\rm BH}$ of red quasars is smaller than that of normal type 1 quasars.
 In other words, red quasars are more active and rapidly growing BHs than normal type 1 quasars,
 which agrees with the scenario that red quasars are in an intermediate stage between
 merger-driven star-forming phase and a visible normal quasar phase.

\acknowledgements
 This work was supported by the Creative Initiative Program of the National Research Foundation of Korea (NRF),
 No. 2008-0060544, funded by the Korea government (MSIP).
 D.K. acknowledges the fellowship support from the grant NRF-2015-Fostering Core Leaders of
 Future Program, No. 2015-000714, funded by the Korean government.
 J.H.W. acknowledges support by the National Research Foundation of Korea to the Center for Galaxy Evolution Research (2010-0027910).
 E.G. acknowledges the generous support of the Cottrell College Award through the Research Corporation for Science Advancement.

 M.I. and D.K. are Visiting Astronomers at the Infrared Telescope Facility,
 which is operated by the University of Hawaii under Cooperative Agreement no. NNX-08AE38A
 with the National Aeronautics and Space Administration, Science Mission Directorate, Planetary Astronomy Program.

\appendix
\section{NIR SPECTRA OF RED QUASARS}
 We provide reduced NIR spectra of five red quasars (0911$+$0143, 1248$+$0531, 1309$+$6042, 1313$+$1453, and 1434$+$0935).
 Table 3 is an example spectrum of 0911$+$0143. The full version of the reduced spectra
 is available in machine readable table form.

\begin{deluxetable}{ccc}
\tabletypesize{\scriptsize}
\tablewidth{0pt}
\tablenum{3}
\tablecaption{Spectrum of 0911$+$0143\label{tbl3}}
\tablehead{
\colhead{$\lambda$}&	\colhead{$f_{\lambda}$}&	\colhead{$f_{\lambda}$ Uncertainty}\\
\colhead{($\rm \AA{}$)}&	\colhead{($\rm{erg~s^{-1}~cm^{-2}~\AA{}^{-1}}$)}&	\colhead{($\rm{erg~s^{-1}~cm^{-2}~\AA{}^{-1}}$)}
}
\startdata
9381&	-3.388E-17&		2.237E-17\\
9384&	-7.517E-15&		1.877E-17\\
9387&	2.428E-17&		2.087E-17\\
9389&	2.050E-17&		2.272E-17\\
9392&	1.747E-17&		1.711E-17\\
9395&	-7.548E-19&		2.203E-17\\
9398&	2.973E-17&		1.968E-17\\
9400&	5.394E-17&		1.918E-17\\
9403&	3.614E-17&		2.249E-17\\
\enddata
\tablecomments{This table displays only a part of the spectrum of 0911$+$0143. The entire spectra of five red quasars
are available as a tar file in the electronic version of the Journal.}
\end{deluxetable}

\end{document}